\documentclass[conference]{IEEEtran}

\usepackage{cite}
\usepackage{amsmath,amssymb,amsfonts}
\usepackage{graphicx}
\usepackage{xcolor}
\usepackage{array}
\usepackage{url}

\def\BibTeX{{\rm B\kern-.05em{\sc i\kern-.025em b}\kern-.08em
    T\kern-.1667em\lower.7ex\hbox{E}\kern-.125emX}}

\begin{document}

\title{Why Agentic Workflows Fail: A Taxonomy Derived from Literature and Empirical Data}

\author{
  \IEEEauthorblockN{Xin Li, Rami Bahsoon, and Tao Chen}
  \IEEEauthorblockA{
    School of Computer Science\\
    University of Birmingham\\
    Birmingham, United Kingdom\\
    x.li.20@bham.ac.uk
  }
}
\maketitle

\begin{abstract}
Agentic workflows combine large language models with planning, memory, tools, execution environments, and iterative feedback. Their failures are difficult to diagnose because the same user-visible symptom may originate from model behavior, workflow orchestration, external tools, infrastructure, or their interaction. This paper presents a source-aware taxonomy of agentic workflow failures and applies it to evidence from academic literature and GitHub. The taxonomy combines source attribution with three empirical dimensions: a concrete failure-type dictionary, workflow-stage localization, and agent-capability localization. Its central structure is a two-axis localization matrix: workflow stages form the horizontal process axis and agent capabilities form the vertical functional axis. This design moves beyond taxonomies that mainly classify final failure manifestations by localizing intermediate process states inside agent operation. Analysis of 397 literature-derived and 1,196 GitHub-derived failure records, followed by a separate temporal hold-out, shows that literature-derived failures emphasize planning, reasoning, learning, perception, and knowledge-related mechanisms, whereas GitHub-derived failures concentrate on execution-stage, actuation-related, tool, API, and infrastructure failures. The dominant hotspot in the collected GitHub evidence is iterative execution and actuation. These source-specific observations, which may also reflect reporting and observability differences, indicate that agentic workflow reliability should be treated as a systems problem rather than only a model-output problem. The taxonomy supports structured diagnosis and engineering triage, and motivates future research on evaluation, root cause analysis, and failure injection.
\end{abstract}


\section{Introduction}
\label{sec:introduction}

Large language model (LLM) applications are increasingly organized as \emph{agentic workflows}: systems in which an LLM interprets a goal, plans subtasks, invokes tools or external services, observes intermediate results, revises its behavior, and produces an outcome. Such workflows now appear in software engineering, web automation, data analysis, research assistance, and multi-agent collaboration. Prior work has studied foundation-model risks, LLM trustworthiness, and governance concerns~\cite{Bommasani2021FoundationModels,Weidinger2022LMRiskTaxonomy,Bender2021StochasticParrots,NIST2023AIRMF,Liu2024TrustworthyLLMs}; tool-augmented and multi-agent architectures~\cite{Yao2023ReAct,Schick2023Toolformer,Qin2023ToolLLM,Wu2023AutoGen,Li2023CAMEL,Hong2024MetaGPT,Qian2024ChatDev}; and benchmarks for web, software, operating-system, and tool-use agents~\cite{Liu2024AgentBench,Zhou2024WebArena,Xie2024OSWorld,Trivedi2024AppWorld,Yao2024TauBench,Lu2025ToolSandbox,Ma2024AgentBoard}. These studies show that LLM agents can act beyond single-turn text generation, but they also expose new reliability problems: a failed workflow may result from goal misunderstanding, incomplete planning, wrong tool selection, invalid parameter binding, unsafe code execution, stale memory, poor recovery, or infrastructure changes outside the model.

Existing taxonomies only partially address this problem. Traditional defect and dependability taxonomies standardize how software anomalies are recorded and analyzed~\cite{Chillarege1992ODC,Avizienis2004DependabilityTaxonomy,IEEE1044-2009}, while LLM risk taxonomies describe harms such as hallucination, bias, privacy leakage, misuse, and brittleness~\cite{Weidinger2022LMRiskTaxonomy,Das2024LLMSecPrivacySurvey,Xi2023LLMAgentSurvey,Guo2024LLMMultiAgentSurvey}. However, agentic workflows combine model reasoning, prompts, memory, tools, execution environments, APIs, and human feedback. A single observed symptom may therefore cross responsibility boundaries: an agent may choose the wrong action, an external tool may fail, or the two may interact in a way that produces the failure. Existing work often identifies \emph{what} went wrong, but not consistently \emph{where} the failure occurred in the workflow, \emph{which capability} was implicated, or \emph{whether} the primary responsibility lies in the agent, the surrounding system, or their interaction.

This paper addresses this gap by developing and applying a source-aware taxonomy of agentic workflow failures. The taxonomy combines source attribution with three empirical analysis dimensions: (1) a dictionary of recurring failure mechanisms, (2) workflow-stage localization, and (3) agent-capability localization. The stage and capability dimensions can also be inspected jointly through a stage--capability matrix. The goal is not to replace existing software or LLM risk taxonomies, but to connect their insights to concrete failure episodes in multi-step agentic workflows. This structure allows the same failure record to be described by its responsibility boundary, mechanism, process location, and implicated capability.

We construct the taxonomy from 397 literature-derived failure records and refine and empirically apply it using 1,196 GitHub-derived failure episodes. This literature--GitHub comparison is central to the study design; observed differences are source-specific visibility patterns and need not imply differences in the underlying population prevalence of failures. Literature provides conceptual coverage of failures discussed in prior research, while GitHub provides engineering evidence from issues, pull requests, discussions, commits, and related artifacts. The formal research questions are introduced in Section~\ref{sec:findings}, where we align the empirical analysis with three taxonomy dimensions: failure type, workflow stage, and agent capability. We compare literature and GitHub evidence on each dimension and then integrate the results using the stage--capability matrix to identify cross-source failure hotspots.
This paper makes three contributions:
\begin{itemize}
    \item \textbf{A source-aware taxonomy.} We introduce a taxonomy for agentic workflow failures that combines responsibility attribution, failure type, workflow stage, and agent capability.
    \item \textbf{A literature--GitHub empirical comparison.} We apply the taxonomy to 397 literature-derived records and 1,196 GitHub-derived episodes, revealing source-specific visibility patterns rather than establishing differences in underlying failure prevalence.
    \item \textbf{Evidence-based diagnostic insights.} We localize the prominent S3 $\times$ D hotspot in public GitHub evidence, test the source contrast with clustered uncertainty estimates, and illustrate how source attribution can guide engineering triage. Failure injection and monitoring remain future applications.
\end{itemize}

The remainder of this paper is organized as follows. Section~\ref{sec:related_work} reviews related work. Section~\ref{sec:method} describes the methodology. Section~\ref{sec:taxonomy} presents the taxonomy. Section~\ref{sec:findings} reports the empirical results. Section~\ref{sec:discussion} discusses implications, Section~\ref{sec:threats} summarizes threats to validity, and Section~\ref{sec:conclusion} concludes the paper and outlines future work.

\section{Related Work}
\label{sec:related_work}

Research on agentic workflow failures builds on software defect taxonomies, LLM risk taxonomies, agent benchmarks, and empirical studies of LLM-supported software engineering, but these lines of work usually analyze defects, model-level risks, benchmark outcomes, or task-specific failures separately.

\textbf{Software and LLM risk taxonomies.}
Traditional software engineering uses taxonomies to standardize anomaly terminology and guide diagnosis. Orthogonal Defect Classification, the dependability taxonomy by Avizienis et al., and IEEE 1044 distinguish faults, errors, failures, and process feedback mechanisms~\cite{Chillarege1992ODC,Avizienis2004DependabilityTaxonomy,IEEE1044-2009}. These foundations remain useful, but agentic workflows distribute responsibility across models, prompts, memory, tools/APIs, orchestration, and human feedback. A parallel stream studies foundation-model and LLM risks, including hallucination, bias, misuse, privacy leakage, brittleness, coordination failure, and harmful social impact~\cite{Weidinger2022LMRiskTaxonomy,Bommasani2021FoundationModels,Bender2021StochasticParrots,NIST2023AIRMF,Liu2024TrustworthyLLMs,Das2024LLMSecPrivacySurvey,Xi2023LLMAgentSurvey,Guo2024LLMMultiAgentSurvey,Amershi2019HumanAIInteraction}. However, these studies usually operate at the level of risks, governance principles, visible harms, or broad defect concepts rather than concrete failure episodes inside multi-step agentic workflows.

\textbf{Agent architectures, benchmarks, and software-engineering evidence.}
Agent frameworks and benchmarks make workflow failures visible through reasoning traces, tool calls, messages, and environment interactions. ReAct, Toolformer, ToolLLM, AgentBench, AutoGen, CAMEL, MetaGPT, and ChatDev connect LLMs to tools, roles, conversations, and collaborative workflows~\cite{Yao2023ReAct,Schick2023Toolformer,Qin2023ToolLLM,Liu2024AgentBench,Wu2023AutoGen,Li2023CAMEL,Hong2024MetaGPT,Qian2024ChatDev}. WebArena, OSWorld, AppWorld, $\tau$-bench, ToolSandbox, AgentBoard, and MAST evaluate agents in web, GUI, tool-use, long-horizon, and multi-agent settings~\cite{Zhou2024WebArena,Xie2024OSWorld,Trivedi2024AppWorld,Yao2024TauBench,Lu2025ToolSandbox,Ma2024AgentBoard,Cemri2025MAST}. Recent critiques of agent evaluation further argue that benchmark accuracy alone can obscure cost, overfitting, inadequate holdout sets, and reproducibility issues~\cite{Kapoor2024AIAgentsMatter}, reinforcing the need for reusable diagnostic views beyond final task success. Empirical software-engineering studies also expose failures through SWE-bench, SWE-agent, OpenHands, Agentless, ChatGPT-based bug fixing, DevGPT and shared developer--ChatGPT conversations, agent-issue benchmarks, and issue-resolution signals~\cite{Jimenez2024SWEbench,Yang2024SWEAgent,Wang2025OpenHands,Xia2025Agentless,Sobania2023ChatGPTBugFixing,Xiao2024DevGPT,Hao2024SharedConversations,Rahardja2025CanAgentsFixAgentIssues,Ehsani2026IssueResolutionSignals}. These works provide realistic evidence, but their analyses are usually tied to a benchmark, framework, or task outcome rather than a reusable process-level localization scheme.

\textbf{Closest work and diagnostic gap.}
The closest line of work is recent failure analysis and attribution for LLM-based agent systems, especially Cemri et al.'s MAST study~\cite{Cemri2025MAST} and recent work on automated failure attribution in multi-agent systems~\cite{Zhang2025WhoWhen}. MAST is highly relevant because it analyzes concrete failure patterns in multi-agent traces, including specification and system-design failures, inter-agent misalignment, and verification or termination failures; failure-attribution studies further ask which agent or step is responsible for a failed trace. Our taxonomy is complementary, but differs in scope and diagnostic objective. These studies mainly focus on multi-agent coordination, trace-level failure modes, or responsible-agent localization, whereas our unit of analysis is the broader \emph{agentic workflow failure episode}, including single-agent tool use, orchestration, APIs, runtime environments, human feedback, infrastructure, and mixed agent--system interactions.

Existing taxonomy families provide important but partial coverage. Software defect and dependability taxonomies distinguish faults, errors, and failures, but do not localize failures inside LLM-agent workflow stages or capabilities. LLM risk taxonomies capture harms such as hallucination, bias, privacy leakage, and brittleness, but usually describe model-level risks or final user-visible outcomes. Agent benchmarks and multi-agent studies expose realistic traces, but their labels are often tied to benchmark outcomes, coordination patterns, or final trace-level manifestations. These schemes therefore do not directly provide a joint view of \emph{who or what is responsible}, \emph{what mechanism occurred}, \emph{where in the workflow it occurred}, and \emph{which agent capability was implicated}.

Our taxonomy addresses this gap through a source~$\times$ type~$\times$ stage~$\times$ capability structure that adds diagnostic information beyond final-symptom classification. For example, an agent that keeps retrying a failed tool call without using the returned error signal can be coded as an agentic self-regulation failure at S3: Iterative Execution and Refinement and E: Self-Regulation, pointing to retry limits, monitoring, and recovery logic. In contrast, when an agent sends a valid request but the external API has changed or is unavailable, the same surface symptom is coded as a non-agentic External API/Tool Failure at S3 and D: Actuation, shifting remediation toward interface versioning, schema validation, and dependency monitoring.

Thus, MAST and related trace-level studies show that LLM-agent systems fail in recurring ways, while our taxonomy adds a broader process-localized layer for heterogeneous agentic workflows. This layer enables direct comparison between literature-derived and GitHub-derived evidence and supports more specific engineering triage: whether a failure is primarily a reasoning problem, an execution problem, a monitoring problem, an external infrastructure problem, or a mixed interaction among these components.

\noindent\textbf{Coverage comparison with MAST.}
To probe the distinct scope of our taxonomy, we mapped its 65 operational types against the 14 failure modes in MAST~\cite{Cemri2025MAST}. Under this mapping, 63\% of GitHub episodes and 67\% of literature records fall in types without a MAST counterpart. A complementary item-level inspection of 50 randomly sampled episodes judged 39/50 (78\%) not expressible in MAST, compared with 33/50 (66\%) suggested by the type-level mapping for those same episodes; the judgments agreed for 36/50 items. Conversely, all 14 MAST modes can be associated with at least one type in our taxonomy, while MAST provides finer distinctions for inter-agent coordination. These comparisons concern expressibility under the chosen mapping, not superior performance of one taxonomy on every diagnostic task.

\section{Methodology}
\label{sec:method}

\begin{figure*}[t]
\centering
\includegraphics[width=0.98\textwidth]{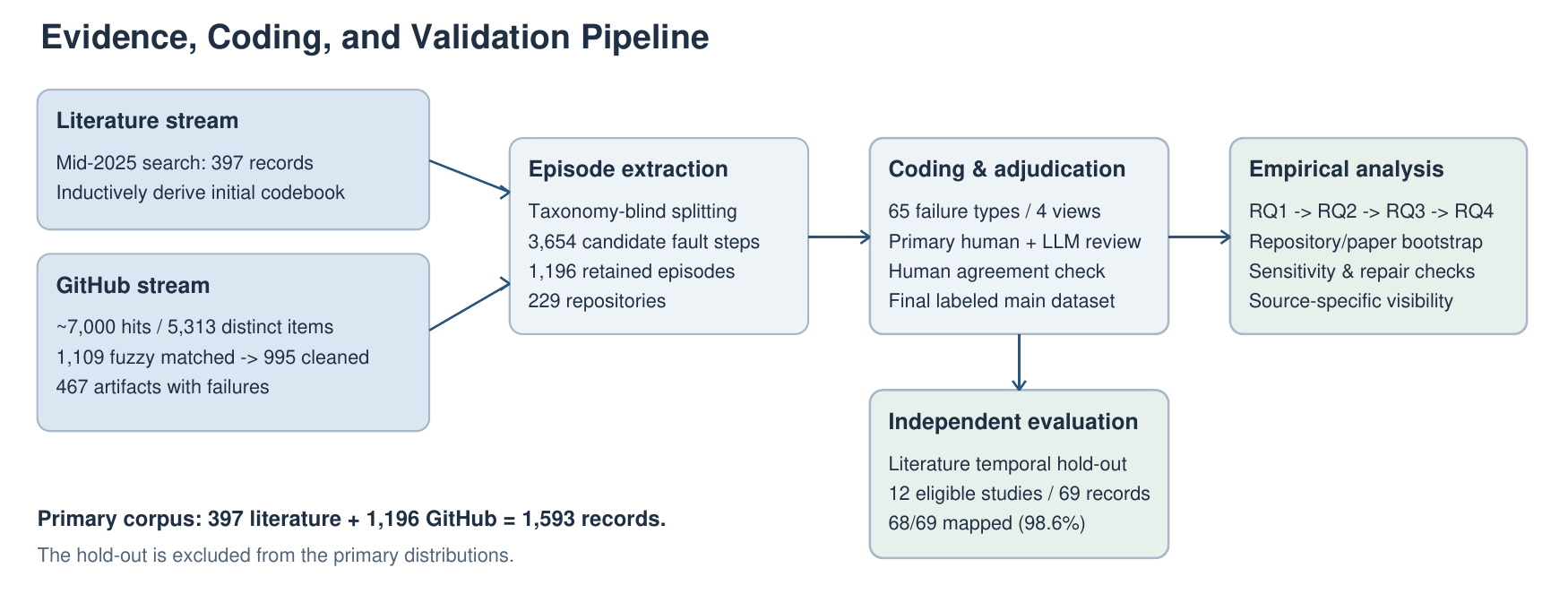}
\caption{Overview of the methodology. Literature evidence is used to derive the initial taxonomy, GitHub evidence is used to extend and stress-test it, and reliability controls are used to adjudicate the final coded dataset.}
\label{fig:method_overview}
\end{figure*}

This study follows a source-aware mixed-method design. Literature evidence captures failure concepts studied in prior work, whereas GitHub evidence captures failures observed when agentic workflow systems are built, integrated, and maintained. Both streams are transformed into the same unit of analysis---a \emph{failure episode}---and then coded with the taxonomy defined in Section~\ref{sec:taxonomy}. Figure~\ref{fig:method_overview} summarizes the workflow.

\subsection{Evidence Collection Protocol}
\label{sec:data_collection}

We used two complementary evidence streams. A mid-2025 literature search covered the preceding five years of work on LLM-based, tool-using, autonomous, and multi-agent workflows. The GitHub stream captured public engineering failures involving tools, APIs, execution environments, user inputs, and infrastructure. The replication package contains the cleaned data, coding prompts, and analysis materials.

\subsubsection{Literature Retrieval and Screening}
\label{sec:lit_protocol}

The literature stream also referred to snowballing practices in software engineering~\cite{kitchenham2007guidelines,wohlin2014guidelines}. Table~\ref{tab:lit_protocol} gives the auditable retrieval protocol. We used broad retrieval terms to avoid missing studies that describe agent failures using different terminology; precision was enforced during screening.

\begin{table*}[t]
\caption{Literature retrieval and screening protocol.}
\label{tab:lit_protocol}
\centering
\footnotesize
\setlength{\tabcolsep}{4pt}
\renewcommand{\arraystretch}{1.12}
\begin{tabular}{@{}p{0.17\textwidth}p{0.77\textwidth}@{}}
\hline
\textbf{Protocol item} & \textbf{Specification} \\
\hline

Databases and expansion & IEEE Xplore, ACM Digital Library, and arXiv; backward and forward snowballing from relevant studies. \\
Search string & (\emph{agent} OR \emph{agentic} OR \emph{LLM agent} OR \emph{autonomous agent} OR \emph{tool-use agent} OR \emph{multi-agent} OR \emph{workflow orchestration} OR \emph{reflection}) AND (\emph{failure} OR \emph{fault} OR \emph{error} OR \emph{bug} OR \emph{breakdown} OR \emph{invalid output}). \\
Inclusion criteria & I1: studies or evaluates an agentic, tool-using, or multi-agent workflow; I2: reports failure phenomena, causes, symptoms, error patterns, or evaluation breakdowns; I3: provides evidence concrete enough to extract at least one failure episode. \\
Exclusion criteria & E1: traditional software defects without agentic workflow behavior; E2: pure model risks without workflow or tool-use context; E3: method-only papers with no identifiable failure evidence; E4: duplicate or superseded versions; E5: inaccessible full text. \\
Screening procedure & Title/abstract screening followed by full-text screening. Ambiguous records were retained until full-text inspection to avoid premature exclusion. \\
\hline
\end{tabular}
\end{table*}

\subsubsection{GitHub Retrieval and Cleaning}
\label{sec:github_protocol}

GitHub search hits were retrieved using API queries and traced to public issues and pull requests, including their discussions, commits, logs, and fix descriptions. Retrieval combined the system-context terms summarized in Figure~\ref{fig:method_overview} with failure-symptom terms such as \textit{failure}, \textit{bug}, \textit{error}, \textit{loop}, \textit{timeout}, \textit{wrong tool}, \textit{invalid output}, and \textit{crash}. The \textit{agentic AI}, \textit{agentic AI fault}, and \textit{agentic system} query families covered ten years ending in May 2025; \textit{agentic workflow} was retrieved through August 2025. To reduce missed records caused by large result sets and API pagination limits, queries were time-sliced into three-month windows from mid-2023 onward and six-month windows before that.

Cleaning used fuzzy keyword filtering, regular expressions, and manual evidence inspection. We removed exact duplicates, inaccessible records, traditional software defects without agentic workflow behavior, unrelated records, news-like items without primary GitHub evidence, automatically generated comments, and records lacking enough context for episode extraction. The retrieval and reduction funnel was: approximately 7,000 search hits (5,313 distinct issues/PRs), 1,109 after fuzzy matching (partial ratio $\geq80$), and 995 after cleaning. Of these 995 candidate artifacts, 467 contained at least one qualifying failure after step-level inspection. The 467 artifacts comprised 211 issues, 207 pull requests, and 49 items subsequently deleted or moved. These 467 artifacts yielded 3,654 candidate single-fault steps and 1,196 qualifying failure episodes for annotation, originating from 229 repositories. The median repository contributed three episodes, and the largest repository accounted for 8.9\% of episodes.

\subsection{Episode Extraction and Taxonomy Construction}
\label{sec:taxonomy_development_method}

A failure episode is a single-fault evidence fragment describing one failure occurrence together with its symptom, context, and, when available, trigger, responsible component, or fix. We use episodes rather than whole papers, issues, or pull requests because one source can contain multiple failures, and several comments can repeatedly describe the same failure. Before any taxonomy labels were assigned, an LLM step-splitting prompt without access to the taxonomy separated each candidate issue/PR into single-fault steps. Episode boundaries were then manually inspected. Distinct fault events were kept separate, while repeated descriptions of the same event were merged. Of the 467 issues/PRs yielding qualifying failures, 301 generated multiple episodes; these within-artifact dependencies motivate clustered inference.

We constructed the taxonomy through a traceable inductive--refinement process rather than by adopting a fixed error list. Literature records were first open-coded to identify recurring failure mechanisms. Candidate labels were consolidated using explicit merge, split, and retention rules: synonymous labels were merged when they referred to the same mechanism; surface-similar failures were split when they differed in cause, diagnostic signal, or remediation strategy; and categories were retained only when they provided diagnostic or engineering value beyond existing labels. This produced an initial literature-grounded codebook containing failure names, definitions, boundary notes, and examples.

The initial codebook was then applied to GitHub episodes as an external stress test against engineering evidence. Episodes that matched an existing definition were coded directly, whereas unmatched episodes were collected as candidate refinements. New types were added only when multiple GitHub records showed a recurring mechanism that could not be reduced to an existing category. After each addition, rename, split, or merge, previously coded records were revisited so that the final dataset used a consistent version of the codebook. The taxonomy was frozen only after this back-coding pass no longer revealed recurring unmatched mechanisms.

The final taxonomy should therefore be read as the stable endpoint of this systematic induction and stress-testing process. The displayed codebook contains 65 operational types within 16 families and retains the AF, NF, and MF source-attribution branches. Counts in the full classification tree are expressed as literature + GitHub and reflect the latest type-level assignments; top-level literature attribution counts include auxiliary labels and should not be treated as a fully human-adjudicated distribution. It has three linked layers. Each GitHub-derived episode receives: (1) a source attribution label, distinguishing Agentic, Non-Agentic, and Mixed failures; (2) the most specific failure type supported by the evidence; and (3) a workflow-stage and agent-capability localization. The source label captures responsibility, the failure type captures the concrete mechanism, and the stage--capability pair captures the intermediate process state in which the failure is introduced or most directly manifested. Section~\ref{sec:taxonomy} defines these labels and their boundary rules.

\subsection{Annotation and Reliability Controls}
\label{sec:annotation_controls_method}

The annotation protocol was designed to avoid forcing ambiguous evidence into overly precise labels. Annotators first determined whether the evidence described a genuine failure and then coded source attribution, failure type, workflow stage, and capability. The primary human annotator held B.Sc. and M.Sc. degrees in computer science and had five years of professional software-engineering experience. DeepSeek-V3.1 provided an auxiliary full-data review through the API model \texttt{deepseek-chat}, with temperature zero and a three-run majority vote (accessed September 18, 2025), using structured prompts included in the replication package. Keyword matching, dictionary matching, regular-expression checks, and structured LLM suggestions were used only to propose candidate labels; final statistics use adjudicated annotations. Disagreements were reviewed against the original evidence and boundary rules. Because one reviewer in the full double-coding process was an LLM-based auxiliary reviewer, we report these values as \emph{human--LLM agreement} rather than as human--human inter-rater reliability. Cohen's $\kappa$ was computed before adjudication, with interpretation guided by established agreement scales~\cite{landis1977measurement}.

We also conducted pilot calibration before final coding. Initial agreement was moderate ($\kappa\approx0.3$--$0.4$), largely because of ambiguous boundary cases and inconsistent interpretations of fine-grained failure types and stage--capability localization. We refined the codebook, evidence-extraction templates, and boundary rules, then froze the protocol before full double coding and adjudication. For the literature stream, human--LLM agreement was $\kappa=0.7211$ for \texttt{Failure\_name}. For the GitHub stream, human--LLM agreement was $\kappa=0.8078$ for the corresponding failure-name labels.

To provide a human-only reliability check across all label dimensions, a second annotator with 14 years of computer-science experience was independently calibrated on a disjoint set and then annotated, blind to previous labels and LLM outputs, 150 GitHub failure episodes, 50 non-failure steps, and 80 literature records. Cohen's $\kappa$ was computed against the original annotations: for GitHub, source attribution 0.71, stage 0.69, capability 0.67, and failure type 0.65; for literature, source 0.79, stage 0.61, capability 0.63, and type 0.72. The literature source agreement is a sampled reliability check and does not convert the 397 literature attribution labels into a fully human-adjudicated distribution. The 50 non-failure steps were also reviewed; no separate classification metric is reported for them. Human--LLM agreement remains auxiliary evidence. Table~\ref{tab:annotation_controls} summarizes the main reliability controls.

\begin{table}[t]
\caption{Independent human--human annotation agreement (Cohen's $\kappa$). The 50 additional non-failure steps are not included in these failure-label rows.}
\label{tab:human_agreement}
\centering
\scriptsize
\setlength{\tabcolsep}{3.0pt}
\begin{tabular}{@{}lrrrrr@{}}
\hline
Dataset & $n$ & Source & Stage & Cap. & Type \\
\hline
GitHub & 150 & 0.71 & 0.69 & 0.67 & 0.65 \\
Literature & 80 & 0.79 & 0.61 & 0.63 & 0.72 \\
\hline
\end{tabular}
\end{table}

\begin{table}[t]
\caption{Annotation and reliability controls.}
\label{tab:annotation_controls}
\centering
\small
\begin{tabular}{@{}p{0.30\columnwidth}p{0.63\columnwidth}@{}}
\hline
\textbf{Control} & \textbf{Implementation} \\
\hline
Source attribution & Classify each episode as Agentic, Non-Agentic, or Mixed according to whether the primary responsibility lies in agent behavior, external conditions, or their interaction. \\
Failure-type coding & Select the most specific dictionary type supported by evidence; use general labels only when finer attribution is unsupported. \\
Matrix localization & Map each episode to one workflow stage and one agent capability to capture its intermediate process state. \\
Tool support & Use fuzzy keyword matching, dictionary matching, regular expressions, and structured LLM suggestions only as candidate labels; final statistics use adjudicated labels. \\
Reliability check & Report full-dataset human--LLM agreement as auxiliary evidence and independent human--human checks on 150 GitHub failures and 80 literature records (with 50 additional non-failure steps reviewed). \\
\hline
\end{tabular}
\end{table}

\subsection{Temporal Hold-Out and Taxonomy Coverage}
\label{sec:temporal_holdout_method}

To evaluate taxonomy coverage beyond the initial mid-2025 retrieval window, we re-ran the original literature search for July 2025--September 24, 2026 under unchanged eligibility criteria. Following deduplication and title/abstract screening, 13 accessible studies entered full-text screening and 12 met the inclusion criteria. The taxonomy was frozen \emph{before} coding these studies; hence this update serves as a temporal hold-out rather than further development data. The 12 studies provided 69 failure records. These records are \emph{not} added to the primary corpus of 397 literature and 1,196 GitHub episodes used for the source-comparison plots and tables.

\subsection{Analysis}
\label{sec:data_analysis_methods}

After adjudication, we tested source contrasts using chi-square tests for overall distributions and two-sided category-versus-rest Fisher exact tests for major categories/cells, with Benjamini--Hochberg correction and Newcombe 95\% confidence intervals for percentage-point gaps. Because failure episodes cluster within repositories and papers, we additionally computed 95\% cluster-bootstrap intervals using the repository and paper, respectively, as resampling units. We examined stability under repository weighting, per-repository caps, exclusion of the top ten repositories, leave-one-repository-out analyses, and collapsing repeated mentions. These checks assess stability of the observed source contrast, not its causal explanation. Then we analyzed the dataset according to the four research questions. RQ1 compares failure-type coverage between literature and GitHub by reporting within-source percentages and percentage-point gaps. RQ2 analyzes the workflow-stage dimension alone, corresponding to the horizontal process axis of the taxonomy. RQ3 analyzes the agent-capability dimension alone, corresponding to the vertical functional axis. RQ4 integrates the preceding results across both evidence sources; the stage--capability matrix is used as a synthesis view of intermediate process states rather than as a separate taxonomy dimension. These analyses are reported in Section~\ref{sec:findings}; broader implications are discussed in Section~\ref{sec:discussion}.

\section{Taxonomy of Agentic Workflow Failures}
\label{sec:taxonomy}

This section defines the taxonomy used in the empirical analysis. It targets agentic workflows and localizes each failure along four views: source attribution, failure type, workflow stage, and agent capability. The central structure is a two-axis stage--capability matrix that localizes intermediate process states inside agent operation.

The taxonomy should be interpreted as the frozen result of the inductive--refinement process described in Section~\ref{sec:taxonomy_development_method}: literature evidence provided the initial open-coded mechanisms, GitHub evidence stress-tested and extended the codebook, and all labels were back-coded after merge, split, rename, or addition decisions.

\noindent\textbf{Design rationale.} Similar symptoms can have different causes. A failed tool call may result from an incorrect agent-generated argument, an unavailable external API, or a malformed tool response combined with missing agent-side validation. Treating all of these as a single ``tool failure'' would obscure both diagnosis and remediation. We therefore separate four coding decisions: \emph{where} the failure occurs in the workflow, \emph{which capability} is implicated, \emph{what mechanism} occurred, and \emph{who or what} is primarily responsible.

\begin{figure*}[!t]
\centering
\includegraphics[width=0.985\textwidth,height=0.82\textheight,keepaspectratio]{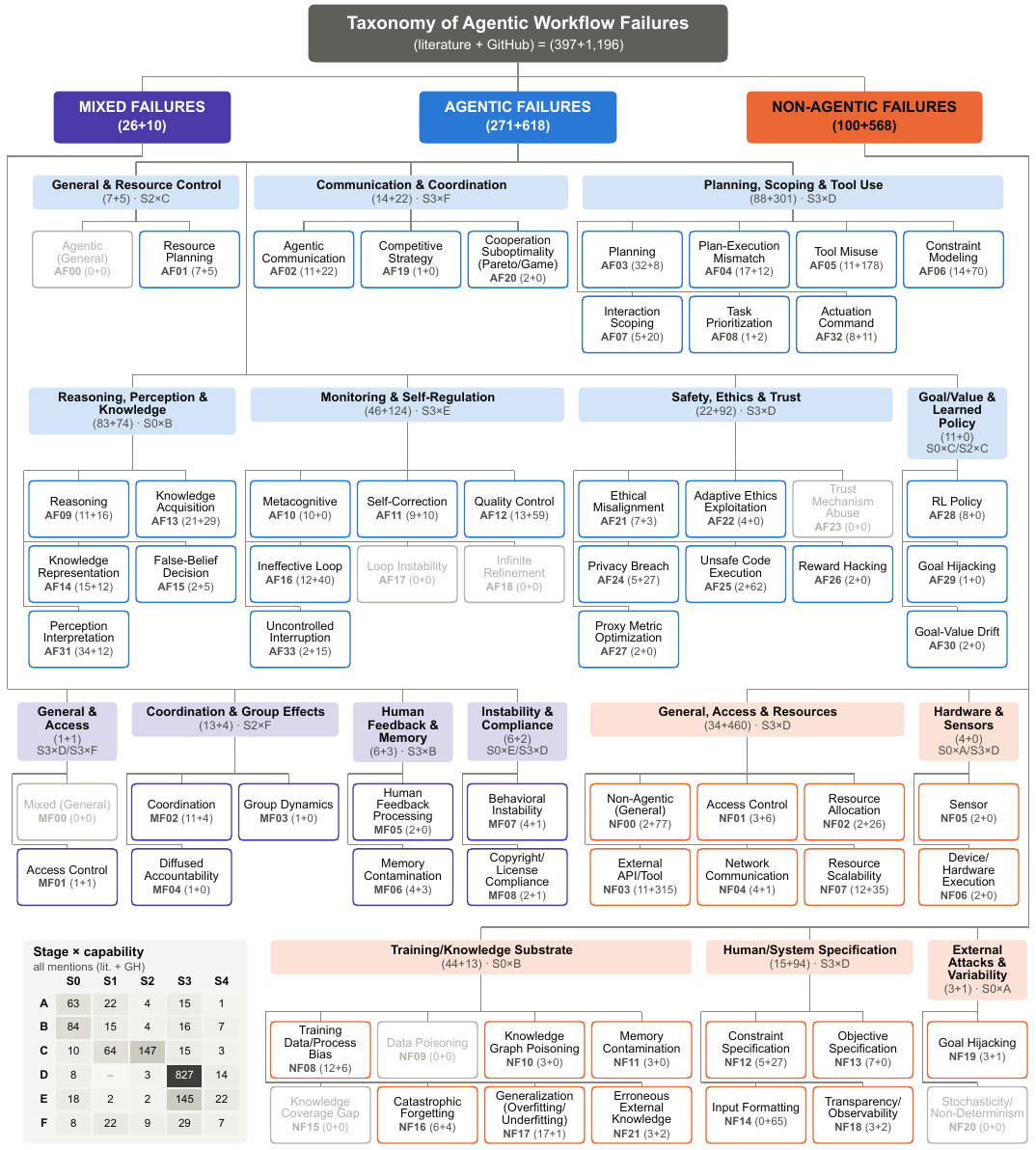}
\caption{Full taxonomy of agentic-workflow failures with three source-attribution categories, 16 failure families, 65 operational failure types, and a stage--capability matrix. Each paired count is (literature + GitHub), totaling 397 + 1,196 failure records. The top-level counts are Agentic (271+618), Non-Agentic (100+568), and Mixed (26+10); literature attribution in these displays includes auxiliary labels and is not a fully human-adjudicated source distribution. Family stage--capability tags describe dominant localization, not an additional type label. Matrix cells show primary stage (S0--S4) by capability (A--F) for 1,586 valid records (397 literature; 1,189 GitHub); the seven GitHub records without complete matrix labels remain in type-level counts. A dash denotes zero; the largest cell is S3 $\times$ D (827).}
\label{fig:agentic_failure_taxonomy}
\end{figure*}

\subsection{Two-Axis Stage--Capability Localization}
\label{sec:taxonomy_two_axis}

Figure~\ref{fig:agentic_failure_taxonomy} displays the complete 65-type dictionary under 16 failure families across the Agentic (AF), Non-Agentic (NF), and Mixed (MF) branches, together with an explicit stage--capability matrix. Each type is labeled with a stable identifier (e.g., AF05 or NF03) and its paired literature--GitHub counts; full operational definitions and decision boundaries are in the replication codebook. The stage axis follows the left-to-right lifecycle of an agentic workflow: S0 captures inherited model and training conditions; S1 captures goal interpretation and scoping; S2 captures strategy and task decomposition; S3 captures iterative execution and refinement; and S4 captures final synthesis and response generation. The capability axis captures the functional capacity implicated by the evidence: A: Perception, B: Learning, C: Reasoning, D: Actuation, E: Self-Regulation, and F: Agent/Human Interaction.

The two axes are intentionally orthogonal. A reasoning failure, for example, may appear during goal interpretation, planning, runtime adaptation, or final synthesis. An actuation failure may appear when choosing a tool route, executing a concrete API call, running code, or formatting the final deliverable. The matrix cells therefore should not be read as separate failure types. They are localization cues for identifying the process state responsible for an observed failure. This is the main distinction between our taxonomy and manifestation-oriented classifications: the same final symptom can be traced to different intermediate workflow states.

\subsection{Source Attribution and Failure Mechanisms}
\label{sec:taxonomy_source_mechanism}

Figure~\ref{fig:agentic_failure_taxonomy} organizes the taxonomy into seven Agentic families, five Non-Agentic families, and four Mixed families, with 34 AF, 22 NF, and nine MF operational codes (including general codes). The chart totals 889 Agentic (271 literature + 618 GitHub), 668 Non-Agentic (100+568), and 36 Mixed (26+10) records. The paired counts represent type-coded evidence, not population incidence: the literature component of top-level source attribution relies partly on auxiliary LLM annotations and is not used to infer a cross-source attribution distribution. Source attribution identifies whether responsibility lies primarily in autonomous agent behavior, external conditions, or their interaction. The concrete mechanisms include AF05 Tool Misuse (11+178), NF03 External API/Tool Failure (11+315), AF31 Perception Interpretation (34+12), and NF14 Input Formatting (0+65). Family-level S$\times$capability labels indicate predominant localization among the family's records, not a restriction to that single matrix cell.

The empirical RQs follow the same taxonomy structure. RQ1 analyzes failure-type coverage and source-normalized gaps. RQ2 analyzes the horizontal workflow-stage dimension. RQ3 analyzes the vertical agent-capability dimension. RQ4 integrates the two evidence sources across these dimensions, using the stage--capability matrix as a synthesis view rather than as an additional taxonomy dimension.

\noindent\textbf{Running example (identical symptom, different source).}
Consider a failed tool invocation during S3: Iterative Execution. If the agent invents an incorrect tool argument, the primary code is Agentic / Tool Misuse / S3 $\times$ D. If the request is valid but an external API is unavailable, it is Non-Agentic / External API/Tool Failure / S3 $\times$ D. If malformed external feedback and omitted agent-side validation jointly contribute, a Mixed source label may be warranted. The symptom is similar in all three cases, but the plausible intervention differs across the agent, interface, and boundary. This example is illustrative, not an extra observed episode.

\subsection{Boundary Rules}
\label{sec:taxonomy_boundary_rules_text}

Each episode receives one \emph{primary} stage and one \emph{primary} capability, rather than an assertion that no other stage or capability contributed. If the introduction point of the responsible defect is evidenced, that point takes precedence; otherwise, annotators use the first observable manifestation. Each episode is coded with the most specific label supported by evidence. General labels are used only when top-level attribution is clear but finer evidence is absent. Mixed labels are reserved for cases where agent behavior and external, human, or system conditions jointly contribute, or where the evidence does not support a single dominant source. Table~\ref{tab:taxonomy_boundary_rules} summarizes the rules used to keep coding consistent.

\begin{table*}[!t]
\caption{Condensed boundary rules for applying the taxonomy.}
\label{tab:taxonomy_boundary_rules}
\centering
\scriptsize
\setlength{\tabcolsep}{3pt}
\renewcommand{\arraystretch}{1.08}
\begin{tabular}{@{}p{0.20\textwidth}p{0.55\textwidth}p{0.19\textwidth}@{}}
\hline
\textbf{Decision point} & \textbf{Rule} & \textbf{Example consequence} \\
\hline
Earliest responsible stage & Prefer the evidenced stage where the defect was introduced; otherwise use its first observable manifestation, not merely the final user-visible symptom. & Goal misunderstanding leading to wrong output is S1; wrong final synthesis after correct work is S4. \\
Stage boundary & Separate task understanding (S1), planning-time choices (S2), runtime action/feedback/recovery (S3), and final response construction (S4). & Bad tool choice in a plan is S2; correct tool with malformed runtime arguments is S3. \\
Capability boundary & Separate decisions and inferences from concrete actions, and separate knowledge/memory from monitoring and recovery. & Wrong algorithm choice is C; malformed shell command is D; context loss is B; repeated failed retry is E. \\
Source attribution & Use Agentic for avoidable agent behavior, Non-Agentic for external conditions outside direct control, and Mixed only when both sides materially contribute or evidence is underdetermined. & API outage is Non-Agentic; wrong arguments are Agentic; malformed tool response plus no validation is Mixed. \\
Specificity & Use the most specific supported failure type; use AF00, NF00, or MF00 only when the source is clear but evidence does not support a finer subtype. & Revisit general labels when logs, comments, or fixes reveal a concrete mechanism. \\
\hline
\end{tabular}
\end{table*}

Together, these rules make the taxonomy process-oriented rather than outcome-oriented. Failure type names the mechanism, source attribution identifies responsibility, and the horizontal--vertical matrix localizes the intermediate workflow state. This separation prevents final symptoms from being overgeneralized while keeping the coding scheme compact enough for empirical comparison.

\section{Empirical Study Results}
\label{sec:findings}

\definecolor{findingbg}{RGB}{247,241,252}
\definecolor{findingborder}{RGB}{122,68,160}
\newcommand{\findingbox}[2]{%
\par\vspace{0.35em}
\noindent\begingroup\setlength{\fboxsep}{5pt}%
\fcolorbox{findingborder}{findingbg}{%
\parbox{\dimexpr\linewidth-2\fboxsep-2\fboxrule\relax}{\small\textbf{#1.} #2}}%
\endgroup\par\vspace{0.45em}}

The primary (non-hold-out) dataset contains 1,593 classified failure records: 397 from literature and 1,196 from GitHub. Seven GitHub records lacked sufficient evidence for a complete stage--capability pair and are excluded only from matrix-level analyses, yielding 1,586 valid matrix records. Percentages are normalized within each source unless otherwise stated. The revised taxonomy tree assigns 618 GitHub records (51.7\%) to Agentic Failures, 568 (47.5\%) to Non-Agentic Failures, and ten (0.8\%) to Mixed Failures. Its paired type-count summaries also show 271, 100, and 26 literature records, respectively; these literature-level source-attribution groupings involve auxiliary LLM annotations and have not been fully human-adjudicated, so we do not treat them as a directly comparable source-attribution distribution.

\subsection{RQ1: Failure-Type Coverage and Source Gaps}
\label{sec:rq1_failure_type_comparison}

\noindent\textbf{Operationalization.}
We compare literature and GitHub records by failure type using
source-normalized percentages. The percentage-point gap is
defined as GitHub percentage minus literature percentage, where
positive values indicate GitHub-dominant types and negative
values indicate literature-dominant types.

\begin{table*}[!t]
\caption{Largest source differences by failure type. Percentages are normalized within each source; gaps are GitHub percentage minus literature percentage.}
\label{tab:failure_type_gap_compact}
\centering
\scriptsize
\begingroup
\setlength{\tabcolsep}{2.2pt}
\renewcommand{\arraystretch}{0.98}
\begin{tabular}{@{}p{0.49\textwidth}@{\hspace{0.02\textwidth}}p{0.49\textwidth}@{}}
\begin{tabular}{|p{0.58\linewidth}|r|r|r|}
\hline
\multicolumn{4}{|c|}{\textbf{GitHub-dominant types}} \\
\hline
\textbf{Failure type} & \textbf{Lit.\%} & \textbf{GH\%} & \textbf{Gap} \\
\hline
External API/Tool Failure (NF03) & 2.8 & 26.3 & +23.6 \\
Tool Misuse Failure (AF05) & 2.8 & 14.9 & +12.1 \\
Non-Agentic Failure (General, NF00) & 0.5 & 6.4 & +5.9 \\
Input Formatting Failure (NF14) & 0.0 & 5.4 & +5.4 \\
Unsafe Code Execution (AF25) & 0.5 & 5.2 & +4.7 \\
Constraint Modeling Failure (AF06) & 3.5 & 5.9 & +2.3 \\
\hline
\end{tabular}
&
\begin{tabular}{|p{0.58\linewidth}|r|r|r|}
\hline
\multicolumn{4}{|c|}{\textbf{Literature-dominant types}} \\
\hline
\textbf{Failure type} & \textbf{Lit.\%} & \textbf{GH\%} & \textbf{Gap} \\
\hline
Perception Interpretation Failure (AF31) & 8.6 & 1.0 & -7.6 \\
Planning Failure (AF03) & 8.1 & 0.7 & -7.4 \\
Generalization Failure (NF17) & 4.3 & 0.1 & -4.2 \\
Plan--Execution Mismatch (AF04) & 4.3 & 1.0 & -3.3 \\
Knowledge Acquisition Failure (AF13) & 5.3 & 2.4 & -2.9 \\
Knowledge Representation (AF14) & 3.8 & 1.0 & -2.8 \\
\hline
\end{tabular}
\end{tabular}
\endgroup
\vspace{-0.5em}
\end{table*}

\noindent\textbf{Results.} Table~\ref{tab:failure_type_gap_compact} shows source-specific concentration at the failure-type level using the names, codes, and paired counts in Figure~\ref{fig:agentic_failure_taxonomy}. External API/Tool Failure (NF03) is the largest GitHub-dominant type: 315 GitHub records (26.3\%) versus 11 literature records (2.8\%), a +23.6 percentage-point gap. Tool Misuse Failure (AF05) and Non-Agentic Failure (General, NF00) have +12.1 and +5.9 percentage-point gaps. The three largest combined types are NF03 (326 records, 20.5\%), AF05 (189 records, 11.9\%), and Constraint Modeling Failure (AF06; 84 records, 5.3\%); together they account for 599 of 1,593 classified records (37.6\%).

\findingbox{Finding 1}{GitHub-derived evidence has a pronounced concentration in tool/API, input-formatting, code-execution, and other operational failure types.}

The three most frequent GitHub types---NF03, AF05, and NF00---account for 570 of 1,196 episodes (47.7\%), whereas the same types account for 24 of 397 literature records (6.0\%). Input Formatting Failure (NF14; 0 literature + 65 GitHub) and Unsafe Code Execution Failure (AF25; 2+62) reinforce this source-specific visibility of operational failure mechanisms.

\findingbox{Finding 2}{Literature-derived evidence places proportionally greater emphasis on perception interpretation, planning, generalization, plan--execution mismatch, knowledge acquisition, and knowledge representation.}

The largest negative percentage-point gap belongs to Perception Interpretation Failure (AF31), with 34 literature records (8.6\%) and 12 GitHub episodes (1.0\%), a -7.6 pp gap. Planning Failure (AF03) follows with 32 literature records (8.1\%) and eight GitHub episodes (0.7\%), a -7.4 pp gap. Generalization Failure (NF17; 17+1), Plan--Execution Mismatch Failure (AF04; 17+12), Knowledge Acquisition Failure (AF13; 21+29), and Knowledge Representation Failure (AF14; 15+12) also have literature-dominant normalized shares. Together, these six types represent 136 of 397 literature records (34.3\%). Although the literature share is higher, AF13 is also observed in 29 GitHub episodes (2.4\% of that source).

These source-specific distributions motivate the separate comparisons of primary workflow stage and agent capability. They describe the observable evidence in the two sampled collections, not a causal difference in true deployment prevalence.

\subsection{RQ2: Workflow-Stage Localization}
\label{sec:rq2_stage_comparison}

\begin{figure}[!t]
\centering
\begin{minipage}[t]{0.49\columnwidth}
    \centering
    \includegraphics[width=\linewidth]{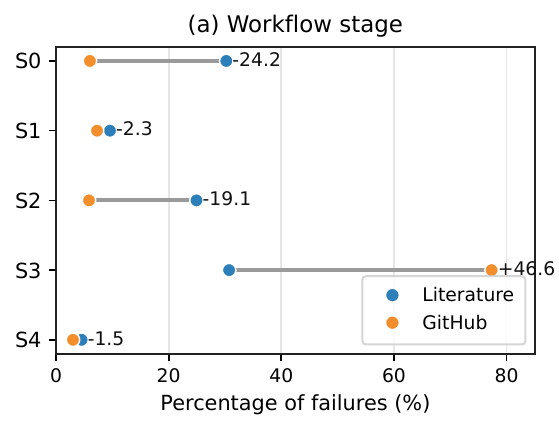}
\end{minipage}
\hfill
\begin{minipage}[t]{0.49\columnwidth}
    \centering
    \includegraphics[width=\linewidth]{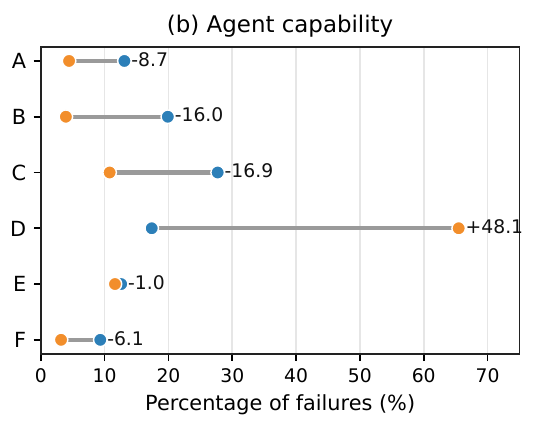}
\end{minipage}
\vspace{-0.5em}
\caption{Comparison of literature-derived and GitHub-derived failures across workflow stages and agent capabilities. Labels indicate the GitHub-minus-literature percentage-point gap. The main marginal gaps are statistically reliable: the stage distribution differs by source, ${\chi^2}(4)=348.92$, $p<0.001$, $V=0.469$, with S3 showing a +46.6 pp gap, 95\% CI [+39.4,+53.2], adjusted Fisher $p<0.001$; the capability distribution also differs, ${\chi^2}(5)=334.19$, $p<0.001$, $V=0.458$, with D: Actuation showing a +48.1 pp gap, 95\% CI [+41.3,+54.1], adjusted Fisher $p<0.001$.}
\label{fig:stage_capability_gap}
\vspace{-0.8em}
\end{figure}

\noindent\textbf{Operationalization.}
We compare the two sources across workflow stages S0--S4 using source-normalized percentages. The stage gap is computed as the GitHub percentage minus the literature percentage and is reported in Figure~\ref{fig:stage_capability_gap}(a).

\noindent\textbf{Results.} Figure~\ref{fig:stage_capability_gap}(a) shows that S3: Iterative Execution and Refinement has the largest positive source gap. S3 contains 925 GitHub records (77.3\%) and 122 literature records (30.7\%), yielding a +46.6 percentage-point GitHub gap. In the combined dataset, S3 contains 1,047 records (65.7\%), making it the dominant workflow stage overall.

\findingbox{Finding 3}{S3: Iterative Execution and Refinement is the dominant GitHub stage and the largest combined-stage hotspot, while literature-derived evidence is more evenly distributed across S0, S2, and S3.}

The dominance of S3 indicates that many observed failures are localized after the workflow has moved beyond problem interpretation and planning. However, the difference between 77.3\% in GitHub and 30.7\% in literature shows that practical failure evidence is especially concentrated in runtime execution, iterative refinement, feedback handling, and environment-facing action. This pattern is consistent with the high frequency of tool, API, input-format, code-execution, and loop-related failure types reported in RQ1.

Across the four GitHub retrieval query families, S3 accounts for 51--89\% of records (versus 30.7\% for the original literature corpus), indicating that the observed concentration is not confined to one query family.

\findingbox{Finding 4}{Literature-derived evidence retains stronger coverage of earlier workflow stages, especially S2: Strategy and Task Decomposition and S0: Foundational Model and Training.}

The largest negative stage gap is S0, with 120 literature records (30.2\%) and 72 GitHub records (6.0\%), yielding a -24.2 percentage-point gap. S2 also has a negative gap: 99 literature records (24.9\%) versus 70 GitHub records (5.9\%), yielding a -19.1 percentage-point gap. S1 and S4 are smaller in both sources; S1 remains relatively close across the two sources, while S4 remains proportionally higher in the literature data than in GitHub.

This distribution shows that the literature data provides broader coverage of pre-execution failures, including model-level behavior, goal interpretation, strategy formation, and task decomposition, whereas these stages form a much smaller share of the practice-oriented GitHub corpus.

\subsection{RQ3: Agent-Capability Localization}
\label{sec:rq3_capability_comparison}

\noindent\textbf{Operationalization.}
We compare literature and GitHub records across six agent capabilities: A: Perception, B: Learning, C: Reasoning, D: Actuation, E: Self-Regulation, and F: Agent/Human Interaction. Percentages are normalized within each source, and source gaps are reported in Figure~\ref{fig:stage_capability_gap}(b).

\noindent\textbf{Results.} Figure~\ref{fig:stage_capability_gap}(b) shows that D: Actuation is the largest GitHub-dominant capability. It contains 783 GitHub records (65.5\%) and 69 literature records (17.4\%), producing a +48.1 percentage-point gap. In the combined dataset, D: Actuation contains 852 records (53.5\%), making it the dominant capability overall.

\findingbox{Finding 5}{GitHub-derived failures are dominated by D: Actuation, indicating that practice-oriented failures are strongly concentrated in the capability responsible for executing actions.}

This result refines the stage-level finding. GitHub failures are not only located in S3; they are specifically concentrated in the capability that performs environment-facing operations. These operations include tool calls, code execution, API invocation, file manipulation, command execution, and integration with external systems. This explains why GitHub-dominant failure types in RQ1 are operational rather than purely cognitive.

The actuation contrast also remains positive under repository-clustered uncertainty estimates, as reported in Section~\ref{sec:robustness_results}.

\findingbox{Finding 6}{Literature-derived failures emphasize C: Reasoning and B: Learning, while E: Self-Regulation remains comparatively similar across the two sources.}

C: Reasoning contains 110 literature records (27.7\%) and 129 GitHub records (10.8\%), producing a -16.9 percentage-point gap. B: Learning contains 79 literature records (19.9\%) and 49 GitHub records (4.1\%). In contrast, E: Self-Regulation accounts for 50 literature records (12.6\%) and 139 GitHub records (11.6\%), producing only a -1.0 percentage-point gap. Thus, monitoring, correction, and iterative control failures are more balanced across the two evidence sources than reasoning- or learning-related failures.

\subsection{RQ4: Integrated Cross-Source Findings}
\label{sec:rq4_matrix_hotspots}

\begin{figure*}[!t]
\centering
\begin{minipage}[t]{0.32\textwidth}
  \centering
  \includegraphics[width=\linewidth]{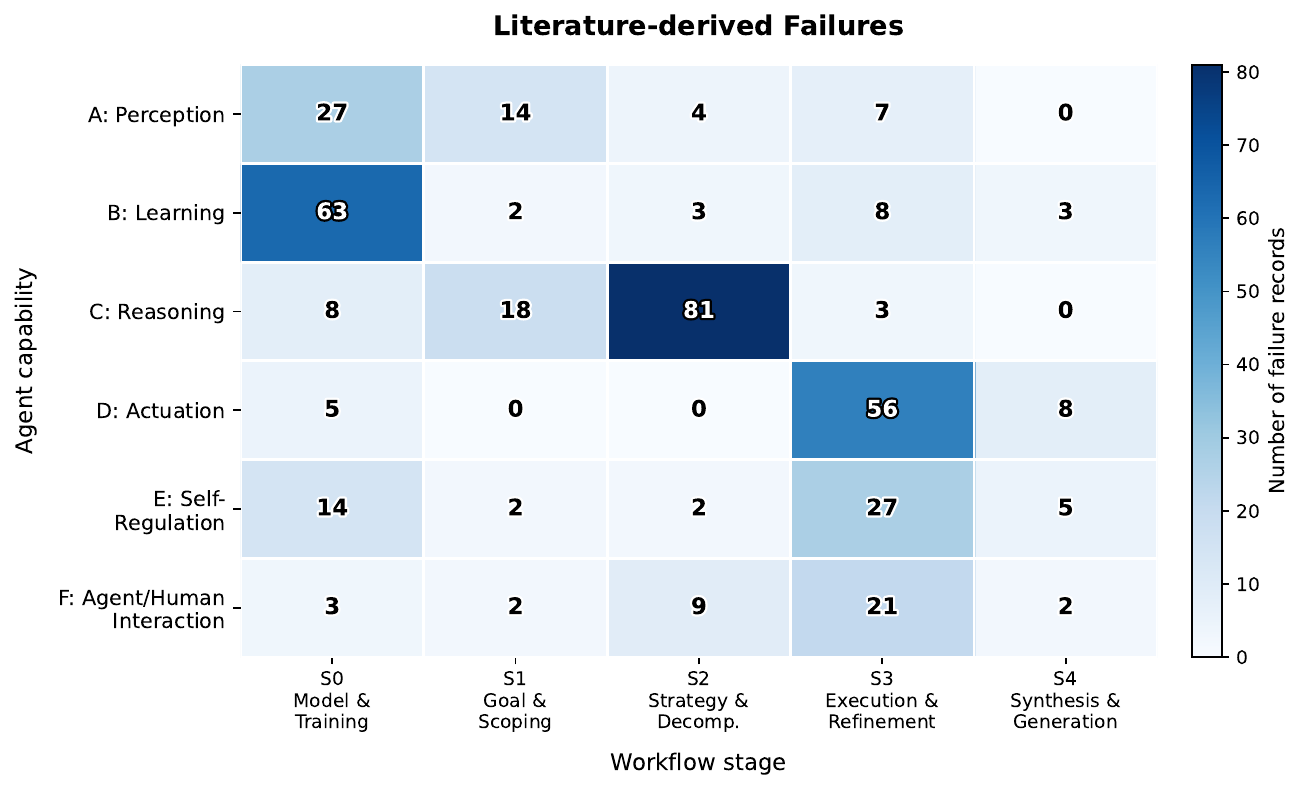}
  \vspace{-0.5em}
  \centerline{\scriptsize Literature}
\end{minipage}\hfill
\begin{minipage}[t]{0.32\textwidth}
  \centering
  \includegraphics[width=\linewidth]{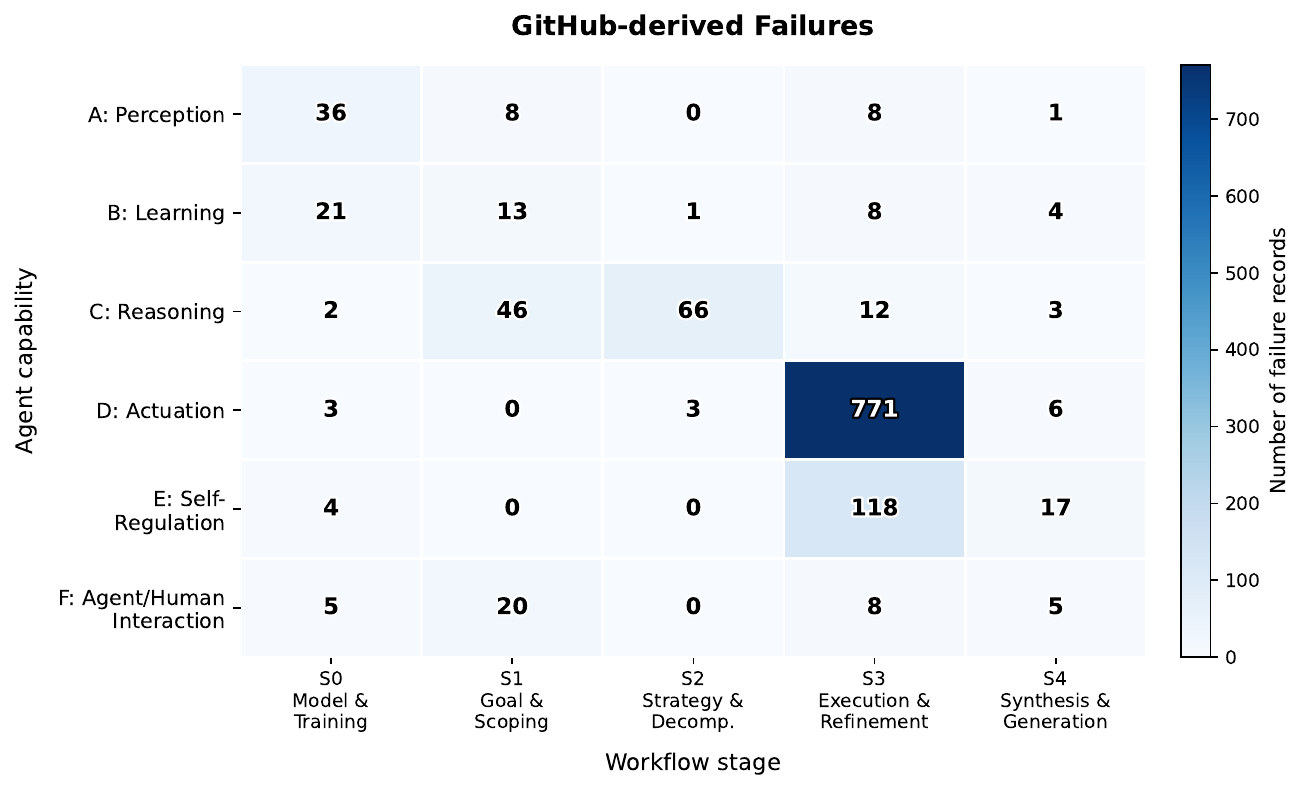}
  \vspace{-0.5em}
  \centerline{\scriptsize GitHub}
\end{minipage}\hfill
\begin{minipage}[t]{0.32\textwidth}
  \centering
  \includegraphics[width=\linewidth]{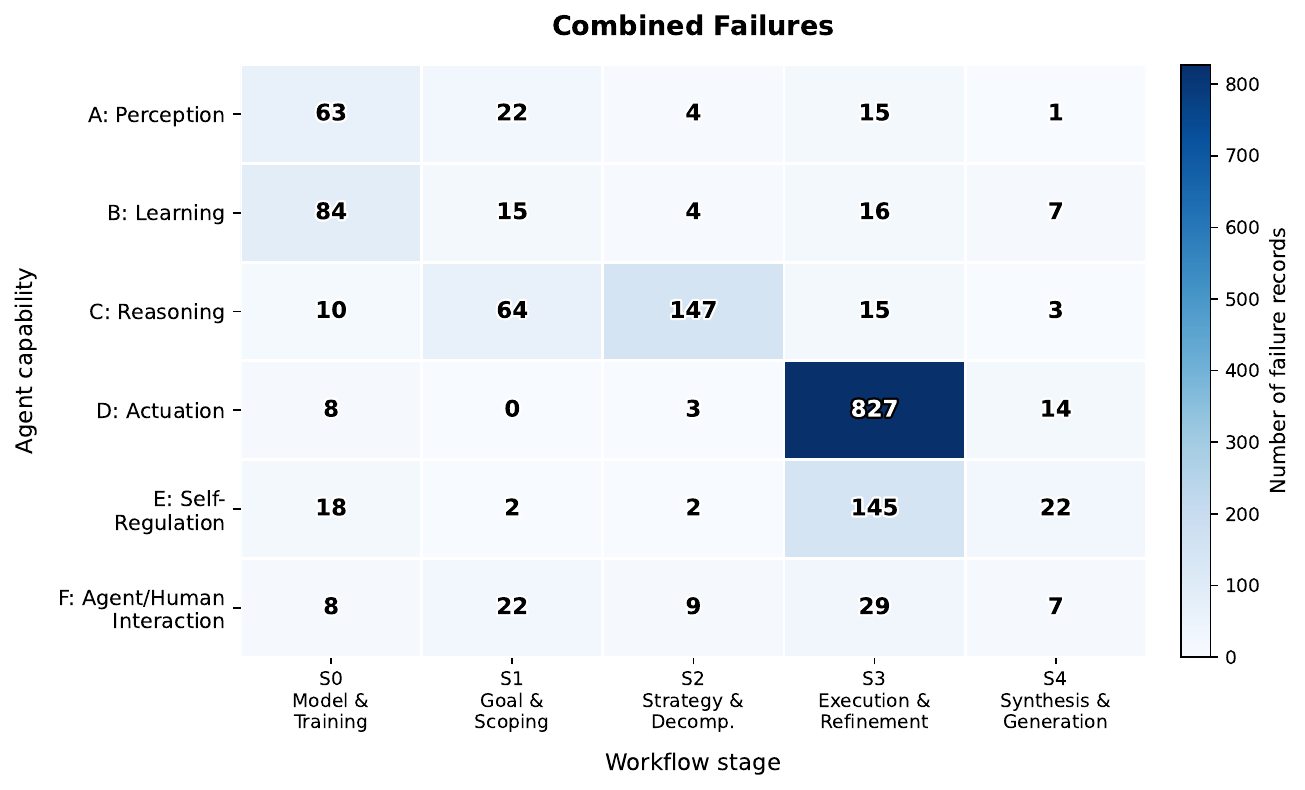}
  \vspace{-0.5em}
  \centerline{\scriptsize Combined}
\end{minipage}
\caption{Stage--capability heatmaps for literature (left), GitHub (center), and the combined dataset (right). Each cell shows the number of failures localized to a workflow stage and capability pair.}
\label{fig:heatmaps}
\end{figure*}

\noindent\textbf{Operationalization.} RQ4 integrates the preceding three analyses rather than introducing a fourth independent taxonomy dimension. We combine the literature and GitHub evidence across failure type, workflow stage, and agent capability to identify the main conclusions supported by both marginal distributions and their intersections. The stage--capability matrix in Figure~\ref{fig:heatmaps} is used as a synthesis view because it shows how the workflow-stage and agent-capability dimensions intersect in each source and in the combined dataset.

\noindent\textbf{Results.} Integrating the three preceding dimensions yields a consistent cross-source conclusion: the failure surface visible in public GitHub evidence is concentrated in runtime execution and actuation, while the literature provides broader coverage of planning-, reasoning-, perception-, and learning-oriented mechanisms. Figure~\ref{fig:heatmaps} shows the literature, GitHub, and combined stage--capability distributions side by side. The largest matrix hotspot contains 827 combined records, corresponding to 52.1\% of the 1,586 records with valid matrix labels. It contains 771 GitHub records and 56 literature records. The next largest cells are S2 $\times$ C: Reasoning (147 records, 9.3\%) and S3 $\times$ E: Self-Regulation (145 records, 9.1\%).

\findingbox{Finding 7}{After integrating the two sources, the dominant empirical conclusion is that S3 $\times$ D: Actuation is the main practical bottleneck, and it is mainly driven by GitHub-derived evidence.}

The integrated result shows that the main practice-oriented cluster is more specific than either ``S3 failures'' or ``actuation failures'' alone. It is the combination of iterative execution and actuation that dominates the GitHub evidence. S3 $\times$ D accounts for 771 of 1,189 valid GitHub matrix records (64.8\%), compared with 56 of 397 literature records (14.1\%). This cell alone is larger than any other stage--capability cell in either source.

This intersection explains why the stage and capability results reinforce each other rather than representing two unrelated patterns.

\findingbox{Finding 8}{The integrated comparison reveals source-specific visibility: literature-derived records emphasize reasoning- and learning-oriented cells, while public GitHub records emphasize execution-time actuation and self-regulation cells.}

The literature heatmap shows stronger secondary concentrations in S2 $\times$ C and S0 $\times$ B, whereas GitHub is dominated by S3 $\times$ D and also contributes heavily to S3 $\times$ E.

The top three integrated cells--S3 $\times$ D, S2 $\times$ C, and S3 $\times$ E--account for 1,119 combined records, or 70.6\% of all valid matrix records. However, they represent different mechanisms and sources: S3 $\times$ D is strongly GitHub-driven, S3 $\times$ E is also GitHub-heavy, and S2 $\times$ C has a larger literature contribution. RQ4 therefore distills the preceding analyses into two research conclusions: the collected GitHub evidence is dominated by execution-time actuation, while the reviewed literature offers stronger coverage of reasoning, learning, and planning mechanisms.

\subsection{Robustness and Additional Diagnostic Checks}
\label{sec:robustness_results}

\noindent\textbf{Clustered source contrasts.}
Resampling papers and repositories as clusters yielded the following GitHub-minus-literature differences with 95\% cluster-bootstrap confidence intervals: S3, +46.6 percentage points [39.1, 53.6]; D: Actuation, +48.1 [40.7, 55.9]; and S3 $\times$ D, +50.7 [43.3, 58.7]. These intervals account for clustering and differ from the record-level Newcombe intervals shown in Figure~\ref{fig:stage_capability_gap}. The S3 difference remained within +43 to +47 points under repository weighting, per-repository caps, exclusion of the ten largest repositories, leave-one-repository-out analysis, and collapsing repeated mentions. When the literature was restricted to 22 empirical LLM-agent studies, differences remained for S3 (+47.2 points; 95\% cluster-bootstrap CI [37.3, 56.5]) and S3 $\times$ D (+52.1 points; [42.5, 61.7]). This establishes a persistent \emph{observed source contrast} under these tests but does not identify whether reporting practices, retrieval, observability, or underlying incidence generate it.

\noindent\textbf{Diagnoses within the main hotspot.}
Among the 771 GitHub episodes in S3 $\times$ D, 61\% were non-agentic and 39\% agentic; the reported breakdown includes 41\% external API/tool failures and 20\% tool misuse. Thus, even within one stage--capability cell, a similar ``failed tool call'' symptom may point to different responsibility boundaries. In an exploratory inspection of 50 episodes, 17 cases had a visible fix. All ten non-agentic cases with visible fixes were addressed outside the agent, whereas four of seven agentic cases were fixed in the agent itself; the remaining three used configuration, framework, or documentation changes. These descriptive results support differentiated triage \emph{in the observed sample}; they do not demonstrate intervention efficacy or generalizable repair success rates.

\noindent\textbf{Temporal hold-out.}
The additional 12 eligible studies from July 2025--September 24, 2026 yielded 69 failure records, of which 68 (98.6\%) could be assigned to existing operational types without expanding the frozen taxonomy. The one unmatched record concerns ``context collapse'' and is a candidate for a new lower-level type; no new top-level family, stage, or capability was required. This is a coverage check on later literature, not a recomputation of the 1,593-record primary dataset (Section~\ref{sec:temporal_holdout_method}).

\section{Discussion}
\label{sec:discussion}

\newcommand{\fref}[1]{\textbf{\emph{Finding~#1}}}

The data support three conclusions about \emph{source-specific visibility}. First, public GitHub artifacts display strong concentrations in execution and actuation, whereas the sampled literature makes other failure locations more visible; these contrasts are not estimates of differences in underlying real-world incident rates. Second, source attribution adds diagnostic information beyond the surface symptom, as shown by the mix of agentic and non-agentic failures inside S3 $\times$ D. Third, robustness checks and the temporal hold-out strengthen confidence in the observed pattern and taxonomy coverage, while leaving the causes of the source difference unresolved.

\subsection{Implications for Academic Research}
\label{sec:discussion_academic}

\noindent\textbf{Visibility rather than proven prevalence.}
\fref{2}, \fref{3}, \fref{4}, and \fref{8} describe where the two public evidence collections differ. Sampling frames, developer reporting incentives, indexing and retrieval, and the observability of intermediate agent decisions differ between research papers and GitHub issues/PRs. Studies of software-repository mining have likewise identified limits of representativeness and reporting~\cite{Kalliamvakou2014Promises,Bird2009FairBalanced}. Clustered and subset analyses reduce some compositional concerns but cannot distinguish an unreported failure from an absent failure. For this reason, we avoid equating the observed source contrast with a causal research--practice gap.

\noindent\textbf{Complementarity with trace-oriented taxonomies.}
Our 65-type to 14-mode mapping and the 50-item inspection suggest additional expressibility for tool/API, runtime, and external-environment failures, while MAST remains more detailed for inter-agent coordination (Section~\ref{sec:related_work}). Broader scope is not itself proof of higher diagnostic accuracy; future studies should compare coder reliability, utility, and repair decisions on common labeled cases.

\noindent\textbf{Evaluation beyond final-task success.}
\fref{3}, \fref{5}, and \fref{7} motivate reporting tool-call validity, execution results, feedback use, retry histories, and recoverability, alongside final answers. The added temporal hold-out shows that most later-literature records fit the frozen codebook, but a context-collapse case suggests that the lower-level failure dictionary should remain open to refinement.

\subsection{Implications for Industrial Deployment}
\label{sec:discussion_industry}

\noindent\textbf{Boundary-aware engineering triage.}
The dominant S3 $\times$ D cell includes both non-agentic (61\%) and agentic (39\%) episodes. These can share a failed-tool-call symptom while calling for different investigations: API contracts, credentials, or dependency health versus tool choice, argument construction, or agent-side validation. The limited visible-fix sample is compatible with this distinction (all 10 non-agentic cases fixed outside the agent; 4 of 7 agentic cases fixed in the agent), but is exploratory and should not be interpreted as a prospective evaluation of a repair method.

\noindent\textbf{What the taxonomy can and cannot yet establish.}
Source-aware coding can organize investigations and make responsibility boundaries explicit. It does not itself supply a root-cause graph, a tested fault-injection procedure, or a deployed monitoring system. Building and evaluating those interventions, and testing whether the labels improve engineering outcomes relative to symptom-only triage, remain future work.

Overall, the evidence favors a systems-oriented view of agentic reliability, tempered by the reporting and observability limitations of the two public evidence sources.

\section{Threats to Validity}
\label{sec:threats}

\noindent\textbf{Construct validity.}
The analysis describes failure \emph{episodes}, not unique projects or independent incidents. A taxonomy-blind LLM split candidate issues/PRs into single-fault steps before coding, and the resulting boundaries were manually inspected, but step splitting remains interpretive. Failures may span several stages or capabilities; the labels denote a primary localization, not exclusive involvement. We prefer the evidenced introduction stage and otherwise the first observable manifestation. We do not estimate the prevalence of multi-stage or multi-capability cases; a stratified sensitivity analysis remains for future work. The fixed codebook also cannot guarantee that every previously unseen lower-level mechanism fits a current type.

\noindent\textbf{Internal validity and annotation reliability.}
Source attribution, failure types, and stage/capability localizations depend on evidence quality and coding judgments. Full-data human--LLM agreement is auxiliary, while the independent human checks cover samples of 150 GitHub failures and 80 literature records. The extra 50 non-failure steps were inspected without a separately reported confusion measure. The literature source-label reliability sample does not make all 397 literature source labels human-adjudicated. Despite explicit boundary rules and adjudication, errors or missed contextual evidence remain possible.

\noindent\textbf{External validity and source visibility.}
The GitHub data represent 1,196 episodes from 467 candidate issues/PRs across 229 repositories rather than a random sample of deployed agents. Public issues expose execution and integration problems more readily than some internal decisions; papers have different selection, reporting, and abstraction conventions. Repository-clustered confidence intervals, repository exclusions, weighting, and the restricted 22-study literature comparison show that the measured contrasts persist across several checks, but cannot distinguish sampling/observability biases from differences in underlying failure incidence. The 69-record temporal hold-out supports codebook coverage of later studies without establishing saturation in future settings. Neither the source shares nor the type frequencies should be read as population prevalence estimates.

\section{Conclusion}
\label{sec:conclusion}

We developed a source-aware agentic-workflow failure taxonomy and applied it to 397 literature-derived records and 1,196 GitHub-derived failure episodes. Its distinct coding views capture responsibility attribution, failure type, workflow stage, and agent capability, with the stage--capability matrix highlighting intermediate process locations.

Public GitHub evidence is concentrated in S3: Iterative Execution and D: Actuation, particularly their intersection, whereas the sampled literature provides greater visibility into planning, learning, reasoning, and other regions. These source contrasts remain under cluster-bootstrap and repository sensitivity analyses, including a restriction to empirical LLM-agent studies, but do not reveal their causal origin. Independent human-label checks strengthen assessment of coding reliability. In a separate temporal hold-out, 68 of 69 later-literature failure records matched existing types.

The exploratory analysis of visible fixes illustrates how source attribution distinguishes agent-side from external-system intervention loci within a shared failure hotspot. Future research should evaluate whether this additional localization improves triage outcomes and develop and test root-cause analysis, failure injection, and runtime monitoring techniques. We therefore view agentic-workflow reliability as a source-aware systems problem rather than solely a model-output problem.

\section*{Data Availability}
\label{sec:data_availability}

The replication package for this study contains the 397 literature-derived and 1,196 GitHub-derived failure records, the taxonomy codebook, annotation labels, coding prompts, and supporting analysis scripts. These materials support inspection of the coding protocol and reproduction of the empirical analyses reported in this paper. The currently available replication package is hosted at \url{https://anonymous.4open.science/r/literature-github-AWfailure-taxonomy}.

\bibliographystyle{IEEEtran}
\bibliography{bib/references}

\end{document}